# Atomic-scale access to long spin coherence in single molecules with scanning force microscopy

**Authors:** Lisanne Sellies[1]*, Raffael Spachtholz[1], Philipp Scheuerer[1], Jascha Repp[1]*

**Affiliations:**

[1]Institute for Experimental and Applied Physics, University of Regensburg, 93040 Regensburg, Germany.

*Correspondence to: lisanne.sellies@ur.de; jascha.repp@ur.de.

**Understanding and controlling decoherence in open quantum systems is of fundamental interest in science, while achieving long coherence times is critical for quantum information processing[1,2]. Although great progress was made for individual, isolated systems, and electron spin resonance (ESR) of single spins with nanoscale resolution has been demonstrated[3-6], the construction of complex solid-state quantum devices requires ultimately controlling the environment down to atomic scales, as potentially enabled by scanning probe microscopy with its atomic and molecular characterization and manipulation capabilities. Consequently, the recent implementation of ESR in scanning tunneling microscopy (STM)[7-10] represents a milestone towards this goal and was quickly followed by the demonstration of coherent oscillations[11,12], tunable dipolar and exchange couplings[13] and access to nuclear spins[14] with real-space atomic resolution. Atomic manipulation[15,16] even fueled the ambition to realize first artificial atomic-scale quantum**

devices[17,18]. However, the current-based sensing inherent to this method limits coherence times[18-20]. Here, we demonstrate pump-probe ESR atomic force microscopy detection of electron spin transitions between non-equilibrium triplet states of individual pentacene molecules. Spectra of these transitions exhibit sub-nanoelectronvolt energy resolution, allowing local discrimination of molecules that only differ in their isotopic configuration. Furthermore, the electron spins can be coherently manipulated over tens of microseconds, likely not limited by the detection method but by the molecular properties. Single-molecule ESR atomic force microscopy can be combined with atomic manipulation and characterization, and thereby paves the way for atomically well-defined quantum components with long coherence times and local quantum-sensing experiments.

**Main Text:** The experimental setup is shown in Fig. 1a. Individual pentacene molecules were adsorbed onto a dedicated support structure to electrically gate the molecule against the tip potential and – at the same time – apply radio-frequency (RF) magnetic fields. This is achieved by a gold microstrip on a mica disc, covered by an insulating NaCl film that is thick enough to prevent electron tunneling between the molecule and the microstrip. A gate voltage $V_G$ was applied to the microstrip to control single-electron tunneling between the molecule and the conductive tip of the AFM[21]. RF magnetic fields were generated from an RF current sent through the microstrip. Experiments were performed at a temperature of 8 K.

To drive and probe ESR transitions, we first bring the closed-shell pentacene molecule to the excited triplet state $T_1$ by driving two tunneling events with pump pulses of $V_G$[22,23]. As shown previously[22], the subsequent decay from $T_1$ into the singlet ground state $S_0$ can be measured by transferring the populations in $T_1$ and $S_0$ after a controlled dwell time $t_D$ to two different charge states. These charge states can then be discriminated in the AFM signal during a probe period, allowing the population decay of $T_1$ to be measured as a function of $t_D$[22].

As seen in Fig. 1b (red curve), this population decay reflects the three lifetimes $\tau_X = 21$ µs, $\tau_Y = 67$ µs and $\tau_Z = 136$ µs of the three triplet states $T_X$, $T_Y$ and $T_Z$, differing from each other significantly. Driving an ESR transition between two of these states by an RF magnetic field of matching frequency effectively equilibrates their populations and thereby strongly affects the overall population decay of $T_1$[24-26]. This is shown in Fig. 1B (black curve), for which the $T_X$-$T_Z$ transition was driven at 1.54 GHz. Around a $t_D$ of 100 µs, the change in the triplet population due to the RF field is maximal for this transition.

To measure an ESR-AFM spectrum of this transition, we therefore fixed $t_D$ to 100.2 µs, and recorded the triplet population as a function of frequency $f_{RF}$ of the driving field. In this single-molecule experiment the pump-probe cycle (Extended Data Fig. 1) was repeated 6400 times in 20 seconds for every data point while the AFM signal $\Delta f$ (see Methods) was measured and time averaged yielding $\langle \Delta f \rangle$. From $\langle \Delta f \rangle$ a dimensionless normalized frequency shift $\Delta f_{norm}$ was derived, which scales linearly with the triplet population; for details see Methods and Extended Data Fig. 2.

Figure 2A shows the resulting ESR signal with an asymmetric shape, which closely resembles the signal shape obtained with optically detected magnetic resonance (ODMR)[24,25]. This asymmetric shape entails information about the nuclear spin system of the molecule; it arises from the hyperfine coupling of the fourteen proton nuclear spins to the electron spins.

The aforementioned data acquisition scheme was optimized for both swiftness and signal-to-noise ratio. We note that – at slower timescales – even every single triplet-to-singlet-decay event can be probed individually[22], providing absolute information about the spin population, as exemplarily demonstrated by the right axis in Fig. 2A.

Also the $T_X$-$T_Y$ transition at 115 MHz can be probed with this technique, see Fig. 2B, being very similar to the one for pentacene in a terphenyl matrix[27]. Note that the smaller ESR signal in this case is due to the smaller difference of the $T_X$ and $T_Y$ decay rates.

These ESR spectra were acquired in absence of a static external magnetic field and therefore probe the zero-field splitting of $T_1$[24,25]. This splitting arises predominantly from the dipole-dipole interaction of the two unpaired electron spins. It is therefore governed by the spatial distribution of the triplet state and can serve as a fingerprint of the molecule. The shift of the $T_X$-$T_Z$ transition frequency by 60 MHz (~ 4% of its value) with respect to pentacene molecules in a host matrix[24,25] can be rationalized by the different environments.

The narrow ESR line (see Fig. 2a) already indicates a long coherence time. To demonstrate the long coherence enabled by our novel detection scheme, we measured Rabi oscillations[28]. The Rabi oscillations shown in Fig. 2c demonstrate that coherent spin manipulation on a μs timescale is possible. In this experiment (Extended Data Fig. 3), the three triplet states were let to decay independently during 45.1 μs (Extended Data Fig. 4), resulting in a strong imbalance of their populations. Subsequently, an RF pulse of variable duration (0 to 7 μs) at resonance with the $T_X$-$T_Z$ transition was applied, driving the population to oscillate between $T_Z$ and $T_X$. During the remaining ~ 50 μs of a fixed total $t_D = 100.2$ μs the triplet states again decay independently from each other, such that after each pulse sequence, predominantly the $T_Z$ population remains and was detected. Note that this Rabi-oscillation measurement scheme gives rise to an overall decaying trend of $\Delta f_{\text{norm}}$ (see blue line in Fig. 2c and Methods).

A decay constant of 2.2 ± 0.3 µs was extracted from the fit in Fig. 2c. Even though a single-molecule experiment avoids ensemble averaging, it still averages over possible fluctuations occurring in time. Here, the nuclear-spin configurations will fluctuate from pump-probe cycle to pump-probe cycle, giving rise to the peculiar line shape reflecting all different nuclear spin configurations (Fig. 2a and b). Similarly, the measured oscillations represent an average over finding the individual molecule in different nuclear spin configurations and, consequently, of a resonance and thus a Rabi frequency differing for every individual pump-probe cycle. Hence, the observed decay of the oscillations is likely dominated by the dephasing from the fluctuating Rabi frequency, limiting the coherence time[29]. Reducing the hyperfine interaction could then increase the coherence time further.

To this end, we studied pentacene-$d_{14}$, that is, fully deuterated pentacene, the ESR spectrum of which is shown in Fig. 3a. Comparing to pentacene-$h_{14}$, the peak shape is similar, but its high-frequency tail is reduced in width by approximately a factor 14. This reduction is due to the smaller hyperfine interaction of deuterium; the width is now likely dominated by nuclear electric quadrupole interaction[27,30]. The left flank of the signal corresponds to a broadening of 0.12 MHz (full width at half maximum, see Methods), being strongly reduced compared to ESR-STM[7]. This sharp onset corresponds to a sub-nanoelectronvolt resolution in energy, enabled by eliminating main decoherence sources. Importantly, no decoherence due to tunneling electrons is induced, since not even a single electron needs to be sent through the molecule during the ESR pulse. Further, the thick NaCl film completely decouples the molecule from the conducting substrate, eliminating decoherence caused by scattering electrons from the latter[11,18,31]. Moreover, in contrast to ESR-STM and magnetic resonance force microscopy[3], the ESR-AFM technique introduced here does not require a magnetic tip. This also avoids decoherence occurring from an interaction between the spin system and the tip's magnetic stray field[7,11,19].

The corresponding Rabi oscillations exhibit a much longer decay time of $16 \pm 4$ μs, see Fig. 3b. This shows that in the above-described experiments on protonated pentacene, decoherence due to the currentless but force-detected ESR, as introduced here, was negligible. Further, it seems likely that for deuterated pentacene the coherence time is still limited by the smaller but nonzero interaction with the nuclear spins and the resulting fluctuating Rabi frequency. Choosing a suitable molecule, ESR-AFM may even allow for coherent electron spin manipulation and read out at millisecond time scales[24,32], while providing atomic-scale real-space control[22,33,34].

We foresee that the combination of single-spin sensitivity and atomic-scale local information allows understanding of the correlation between decoherence phenomena and the local environment. For example, we can locally identify a single pentacene as a different isotopologue from its spectral signature differing distinctly from the vast majority of molecules and image it in its unique environment, as shown in Fig. 3c. Comparing to ODMR data yields a perfect match to a molecule's spectrum that had been identified[35] as pentacene containing one $^{13}$C atom. The hyperfine interaction generally offers a way to manipulate and probe nuclear spins – featuring even longer coherence times[36,37] – *via* the electronic spin system[36,38-41].

The single-molecule ESR-AFM as introduced here opens several novel research directions at once. In contrast to ESR-STM, the system is brought to and studied at out-of-thermal-equilibrium and thereby eliminates the need to measure at (below) liquid-helium temperatures and large magnetic fields. Moreover, not being limited by thermal population imbalance, ESR-AFM enables studying transitions at much smaller energy differences. This, in turn, allows experiments in absence of a static external magnetic field as well as fingerprinting molecules from their zero-field splitting, opening the door to an understanding of how the local environment can modify the latter. The most striking advantage lies in the currentless detection, greatly reducing decoherence arising from the measurement process itself. The increased spin coherence to the ten microseconds scale demonstrated here also comes with sub-nanoelectronvolt spectral and atomic-scale spatial resolution, representing a leap forward for future artificial quantum devices and local quantum-sensing experiments.

**Main references:**

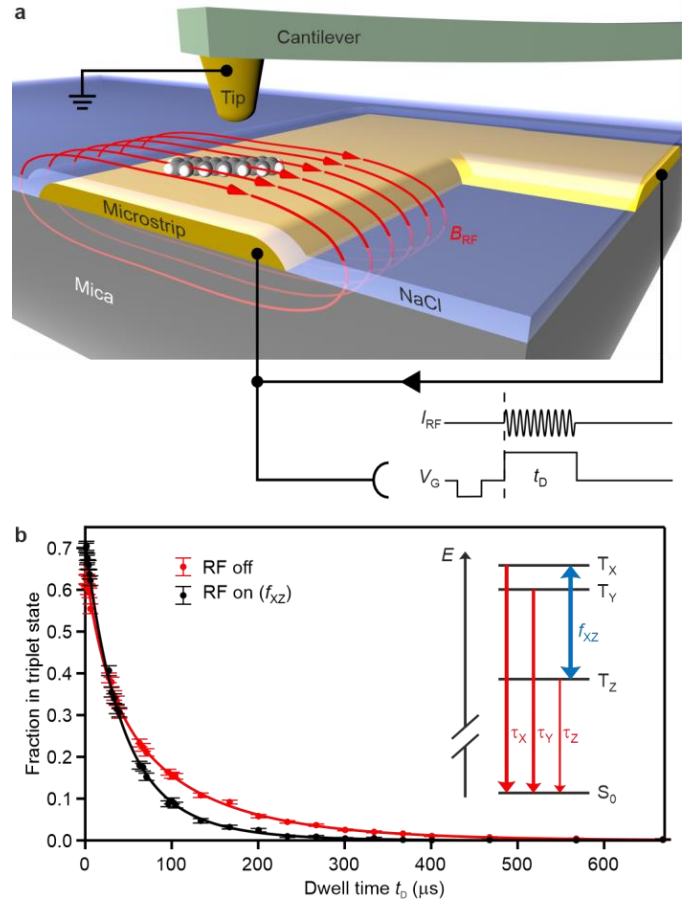

**Fig. 1 | Experimental setup and the effect of resonant driving on the triplet decay. a**, Sketch of the experimental setup. Individual pentacene molecules were adsorbed on an Au(111) microstrip on a mica disc, covered by a NaCl film (> 20 monolayers) preventing electron tunneling between microstrip and molecule. A time-dependent gate voltage $V_G$ was applied to the strip to repeatedly bring the molecule in the triplet excited state $T_1$ by two subsequent tunneling events between molecule and conductive tip. During an experimentally controlled dwell time $t_D$, the molecule can decay to the singlet ground state. A radio-frequency (RF) current $I_{RF}$ was ran through the microstrip to generate an RF magnetic field. After $t_D$ the final state of the molecule was read out as described in the Methods. **b**, Decay of $T_1$ as measured without RF (red) and with

a broadband RF pulse (black). $T_1$ is zero-field split into three states $T_X$, $T_Y$ and $T_Z$ having different lifetimes (inset), such that the RF pulse driving the $T_X$-$T_Z$ transition changes the resulting overall decay. Solid lines represent fits to triple exponential decays. Each data point corresponds to 1920 pump-probe cycles and the error bars were derived from the standard deviation, see ref. 22.

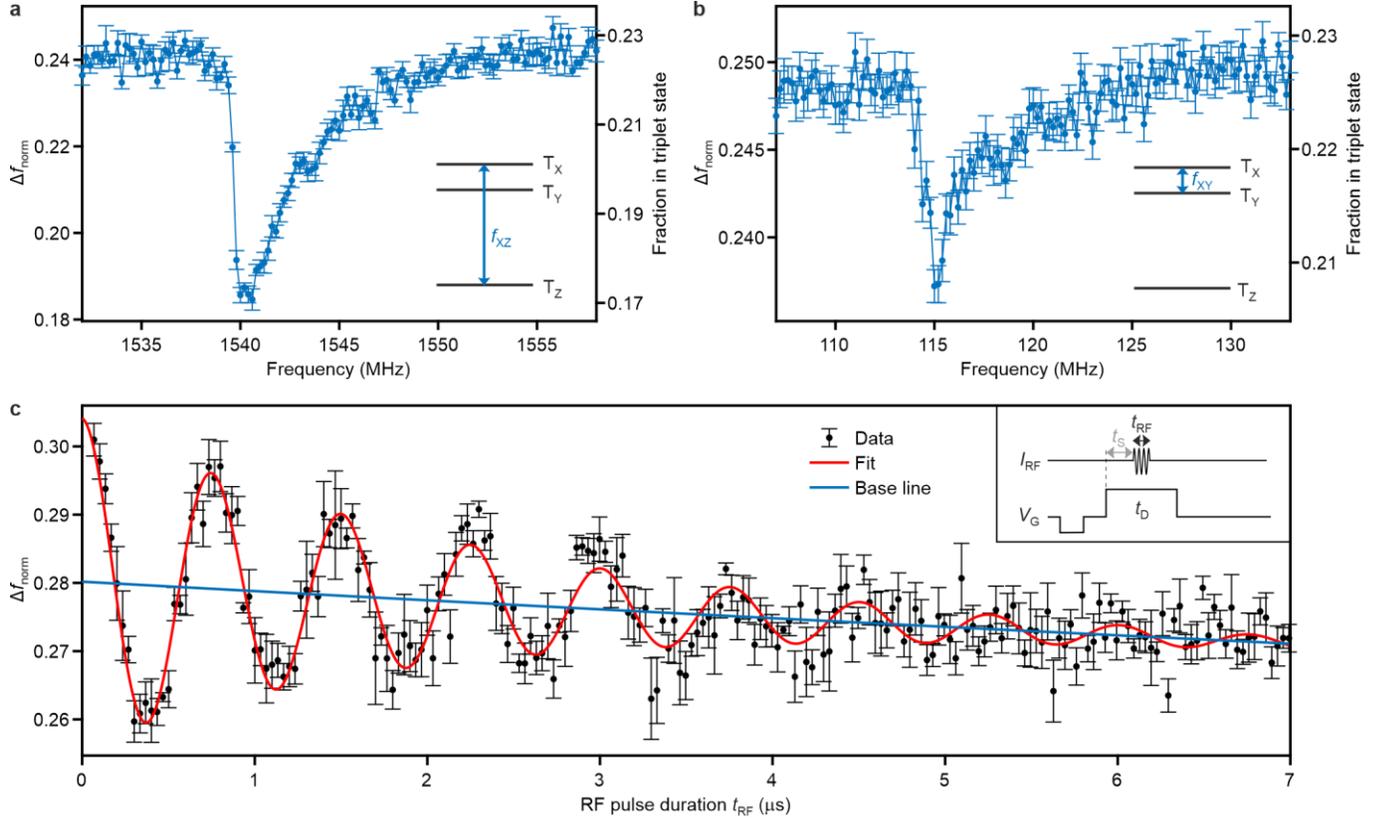

**Fig. 2 | ESR-AFM spectra and Rabi oscillations of a single pentacene molecule. a**, **b**, ESR-AFM spectra of the $T_X$-$T_Z$ and $T_X$-$T_Y$ transitions of a pentacene-$h_{14}$, respectively. The RF frequency was swept at a constant $t_D = 100.2$ μs. The AFM signal $\Delta f$ was normalized to $\Delta f_{\text{norm}}$ as described in the Methods. It can be calibrated against the triplet population[22] at $t_D$, see right axis. The error bars were derived from the standard deviation of repeated measurements. **c**, Rabi oscillations from driving the $T_X$-$T_Z$ transition ($f_{\text{RF}} = 1540.5$ MHz) showing coherent spin manipulation. The pump-probe pulse scheme is shown in the inset ($t_D = 100.2$ μs, $t_S = 45.1$ μs, $t_{\text{RF}}$ variable) and described in the main text. A fit (red line) yields a decay constant of the Rabi oscillations of $2.2 \pm 0.3$ μs.

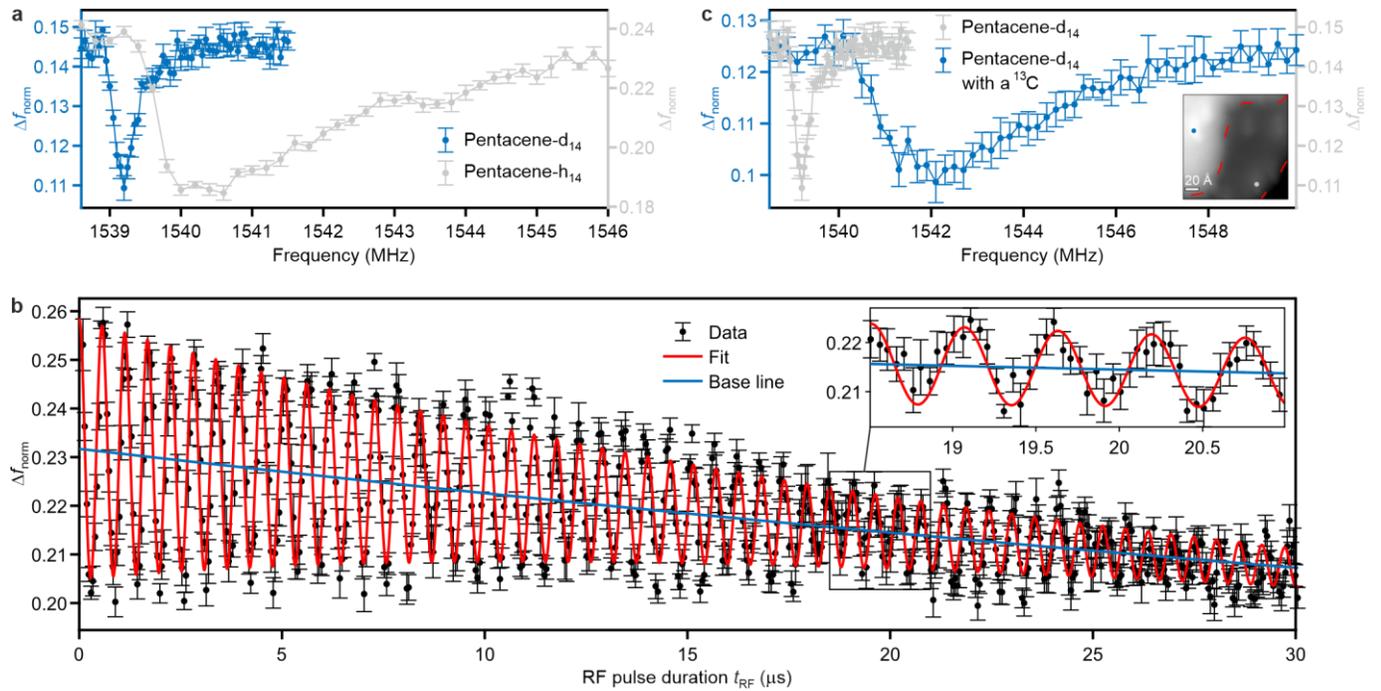

**Fig. 3 | ESR-AFM spectra and Rabi oscillations of different isotopologues of pentacene. a**, ESR-AFM spectrum of pentacene-d$_{14}$ (blue) exhibiting a much narrower resonance in comparison to pentacene-h$_{14}$ (gray). The decreased hyperfine interaction leads to a reduced width of the high-frequency tail. The left flank of the signal corresponds to a broadening of only 0.12 MHz. **b**, The Rabi oscillations of pentacene-d$_{14}$ have a longer decay time of $16 \pm 4$ μs. The pump-probe pulse scheme was the same as the one used for Fig. 2C, but with $t_S = 30$ μs (see Methods). **c**, The high spectral resolution allows identification of different isotopologues of pentacene (blue: pentacene-d$_{14}$ containing a $^{13}$C; gray: pentacene-d$_{14}$) from their spectral signature[35], while AFM permits imaging them in their unique environment (inset, AFM image at constant $\Delta f = -1.4$ Hz and $V_G = 0$ V, the red dashed lines mark the NaCl step edges).

**Methods:**

**Set-up and sample preparation**

Our measurements were performed under ultrahigh vacuum (base pressure, $p < 10^{-10}$ mbar) with a home-built conductive-tip atomic force microscope equipped with a qPlus sensor[42] (resonance frequency, $f_0 = 30.0$ kHz; spring constant, $k \approx 1.8$ kNm$^{-1}$; quality factor, $Q \approx 2.8 \cdot 10^4$) and a conductive Pt-Ir tip. The microscope was operated in frequency-modulation mode, in which the frequency shift $\Delta f$ of the cantilever resonance is measured. The cantilever amplitude was 2 Å (peak-to-peak).

As a sample substrate, we used a cleaved mica disk, on which we deposited gold in a loop structure (diameter, $d = 10.5$ mm; thickness, $t = 300$ nm) by means of electron-beam physical vapor deposition. This gold structure contained a 100 μm wide constriction, where the measurements were performed. A non-conducting spacer material was introduced below the mica disk to prevent eddy-current screening of the radio-frequency (RF) magnetic field. The sample was prepared by short sputtering and annealing cycles (annealing temperature, $T \approx 550$ °C) to obtain a clean Au(111) surface. On half of the sample a thick NaCl film ($> 20$ monolayers) was grown at a sample temperature of approximately 50 °C; the other half of the sample was used for tip preparation, presumably resulting with the tip apex covered with Au. The NaCl film inhibits any electrons to tunnel to or from the gold structure. The voltage that is applied to the gold structure with respect to the tip represents a gate voltage ($V_G$), gating the molecular electronic states against the chemical potential of the conductive tip. The measured pentacene molecules (pentacene-h$_{14}$, Sigma-Aldrich; pentacene-d$_{14}$, Toronto Research Chemicals) were deposited in-situ onto the sample inside the scan head at a temperature of approximately 8 K.

The AC voltage pulses were generated by an arbitrary waveform generator (AWG) (TGA12104, Aim-TTi), combined with the DC voltage, fed to the microscope head by a semi-rigid coaxial high-frequency cable (Coax Japan Co. Ltd.) and applied to the gold structure as $V_G$. The high frequency components of the pulses of $V_G$ lead to spikes in the atomic force microscopy (AFM) signal because of the capacitive coupling between the sample and the sensor. To suppress these spikes, we applied the same pulses with opposite polarity and adjustable magnitude to an electrode that capacitively couples to the sensor.

The RF signal was produced by a software-defined radio (BladeRF 2.0 micro xA4, Nuand), low-pass filtered to eliminate higher frequency components and amplified in two steps (ZX60-P103LN +, Mini-circuits; KU PA BB 005250-2 A, Kuhne electronics). The RF frequency was pulsed using RF switches (HMC190BMS8, Analog devices) which were triggered by the AWG allowing synchronization with $V_G$ and control over the pulse duration. The pulsed RF signal was fed into the microscope head by a semi-rigid coaxial high frequency cable (Coax Japan Co. Ltd.) ending in a loop, inductively coupling the RF signal to the gold loop on the sample. The RF signal transmission of the cables was detected by a magnetic field probe and can be well approximated to be constant over intervals of tens of MHz around the $T_X$-$T_Z$ transition, that is, wider than the spectral features observed in the experiments.

**ESR-AFM pulse sequence**

The description of the measurement of the triplet state lifetime can be found in ref. 22. The ESR-AFM experiments were performed with a similar voltage-pulse sequence, which is shown in Extended Data Fig. 1. Between each individual voltage-pulse sequence, the voltage is set to $V_{\text{deg}}$, the bias voltage, at which the respective ground states of the positively charged ($D_0$) and the neutral ($S_0$) molecule are degenerate. This way, the spin states of the molecule are converted to different charge states and detected[22] by charge resolving AFM[43]. The dwell voltage pulse duration was fixed to 100.2 µs, and simultaneously an RF pulse with a variable frequency was applied. To reduce the statistical uncertainty for a given data-acquisition time, we repeated the pump-probe pulse sequence 320 times per second (instead of 8 times per second for the $T_1$ lifetime measurements). Note that, to prevent the excitation of the cantilever, the durations of the voltage pulses were set to an integer multiple of the cantilever period (33.4 µs). At this high repetition rate of the voltage pulse sequence, the charge states cannot be read out individually. Instead, the AFM signal, that is, the frequency shift $\Delta f$ was averaged over an interval of 20 seconds. This average frequency shift ($\langle \Delta f \rangle$) reflects the ratio of charged and neutral state, and thus the triplet and singlet state, but since the change in $\Delta f$ is very small, it is also sensitive to minor fluctuations in the tip-sample distance. To minimize this dependence on the exact tip height, $\langle \Delta f \rangle$ was normalized using the frequency shift of the charged ($\Delta f^+$) and neutral ($\Delta f^0$) molecule, as: $\Delta f_{\text{norm}} = \frac{\langle \Delta f \rangle - \Delta f^0}{\Delta f^+ - \Delta f^0}$. These frequency shifts were determined at the beginning and end of every 20 seconds data trace (see Extended Data Fig. 2); the charge state was changed by applying small voltage pulses ($V_{\text{set}}^0 = V_{\text{deg}} + 0.3$ V, $V_{\text{set}}^+ = V_{\text{deg}} - 0.3$ V). Tunneling events during the read out of these frequency shifts were minimized by using a tip-sample distance at which the decay constant for the decay of the $D_0$ into the $S_0$ state was around 4 µs. If still a charging event happened, the data trace was discarded. To maximize the rate of the tunneling processes during the voltage pulse sequence, the beginning and end of the voltage pulses were synchronized with the closest turn around point of the cantilever movement.

Note that $\Delta f_{\text{norm}}$ typically deviates from the triplet population, but that for a given measurement a linear relation between them exists. This deviation arises from the voltage pulses that are turned on for 4.3% of the time, during which the frequency shift corresponds to the applied voltages and thus crucially depends on the exact shape of the Kelvin Probe Force parabola[43]. This explains the differences in the base line of the $\Delta f_{\text{norm}}$ signal (without RF or RF off resonance) for different measurements – even for those above the same molecule – due to differences in the position above the molecule. Quantitative results (right axis of Fig. 2A and 2B) can be obtained from a calibration measurement where the population was determined by counting the individual outcomes after each pulse sequence at a repetition rate of 8 per second. This calibration was performed for an RF frequency corresponding to the maximum of the ESR signal, as well as an RF frequency that was off-resonance.

To determine the uncertainty on the ESR-AFM data points, the 20-second data traces were repeated a number of times, and the error bars were extracted as the standard deviation of the mean of these repetitions. Note that the hydrogen spins can have a different configuration for every individual read out[27]. Given our large number of sampling events, we acquire an average over the possible nuclear spin configurations.

**Fitting of the lineshapes**

The shape of the resonance spectra is assumed to be dominated by two contributions. The hyperfine interaction (HFI) together with the fluctuating nuclear spins leads to the peculiar asymmetric line shape. This can be well approximated by a sudden onset at the frequency $f_{\text{onset}}$ followed by an exponential decay of width $f_{\text{decay}}$[44], as

$$\Theta(f - f_{\text{onset}})\exp\left((f - f_{\text{onset}})/f_{\text{decay}}\right),$$

where $\Theta(x)$ denotes the Heaviside function. Secondly, the finite lifetimes of the involved states lead to a lifetime broadening, resulting in a Lorentzian of the form

$$\pi^{-1}\Gamma/((f - f_{\text{res}})^2 + \Gamma^2)$$

centered around each resonance frequency $f_{\text{res}}$ with a full width at half maximum of $2\Gamma$. Accordingly, the experimental resonances are fit to a convolution of the two above functions allowing to extract the broadening due to the HFI and the finite lifetimes separately. We note that non-Markovian processes[29,44] may lead to a deviation from the idealized Lorentzian and that power broadening was avoided in the experiment.

**Rabi oscillations**

The Rabi oscillations were measured using an RF pulse applied around the middle of the dwell voltage pulse with a varying duration and a frequency corresponding to the maximum of the $T_X$-$T_Z$ ESR signal. To illustrate the effect of such an RF pulse, the evolution of the populations of the three triplet states and the singlet state during the dwell voltage pulse were simulated by Maxwell-Bloch simulations, as shown in Extended Data Fig. 3.

The delay time $t_S$, at which the RF pulses started, was fixed for one Rabi-oscillation sweep. The optimal $t_S$ was experimentally determined by sweeping the timing of a π RF pulse over the range of the dwell pulse, as shown for the pentacene-$d_{14}$ molecule in Extended Data Fig. 4. A $t_S > 0$ is needed to initiate an imbalance between the $T_X$ and $T_Z$ states. Similarly, a decay time after the RF pulses is required such that the final triplet population is dominated by only one of these two triplet states. The optimal delay time is, therefore, shortly before the middle of the dwell voltage pulse. Furthermore, it is important that upon increasing the duration of the RF pulse, the sensitivity for differentiating $T_X$ and $T_Z$ does not significantly reduce, otherwise an additional decay of the Rabi oscillations is induced by the read out. Therefore, we chose 30 µs as a delay time for the Rabi oscillations of pentacene-d14, which were probed up to an RF pulse duration of 30 μs.

**Rabi oscillations base line fit**

The base line of the Rabi-oscillation experiment represents the situation of equal populations in the coupled states $T_X$ and $T_Z$ during the pulse – even if the Rabi signal is not decayed yet, it is oscillating around the base line because the populations are oscillating around having equal populations in $T_X$ and $T_Z$. The decay of the base line arises from the decay of the (on average) equally populated $T_X$ and $T_Z$ states into the singlet state during the RF pulse. As the final population of $T_Y$ is independent from the RF signal, it will only give rise to a constant background and will be disregarded in the following.

Hence, the base line is defined by the following: In the initial phase $0 < t < t_S$ all three triplet states decay independently from each other. At the beginning of the RF pulse, that is, at $t = t_S$, the sum of population in $T_X$ and $T_Z$ is

$$P_{XZ}(t_S) = P_0/3(\exp(-k_X t_S) + \exp(-k_Z t_S)),$$

where $k_X = \tau_X^{-1}$ and $k_Z = \tau_Z^{-1}$ are the decay rates of $T_X$ and $T_Z$, respectively, and $P_0$ is the initial total population in the triplet state, such that $P_0/3$ is the initial population in each $T_X$, $T_Y$, and $T_Z$. During the RF pulse, that is, for $t_S < t < t_E$ (with $t_E$ the end of the RF pulse) the RF signal equilibrates (on average) the populations of two of the states, thus at the end of the RF pulse

$$P_{XZ}(t_E) = P_{XZ}(t_S) \exp(-(k_X + k_Z)(t_E - t_S)/2).$$

Finally, for $t_E < t < t_D$ the states decay again independently, giving at the end of the dwell time

$$P_{XZ}(t_D) = P_{XZ}(t_S) \exp(-(k_X + k_Z)(t_E - t_S)/2) \frac{1}{2} \{\exp(-k_X(t_D - t_E)) + \exp(-k_Z(t_D - t_E))\}.$$

which can be rearranged to

$$P_{XZ}(t_D) = P_{XZ}(t_S)\{\exp(-k_X(t_D - t_S))\exp(t_{RF}(k_X - k_Z)/2)$$
$$+ \exp(-k_Z(t_D - t_S))\exp(-t_{RF}(k_X - k_Z)/2)\}/2$$

Note that $P_{XZ}(t_S)$ does not depend on $t_{RF} = t_E - t_S$ and therefore just represents a constant prefactor. The two terms provide contributions to the base line that rise and fall exponentially with $t_{RF}$, respectively. For the specific case and parameters considered here, the prefactor of the rising term is much smaller than the one of the falling term, and is therefore neglected. Since the decay rates were determined (Fig. 1b) for the pentacene-$h_{14}$ molecule, for which the Rabi oscillations were measured, these rates were used for the fitting of the Rabi oscillations of the pentacene-$h_{14}$ molecule (Fig. 2c). In case of pentacene-$d_{14}$ (Fig. 3b), we set $(k_X - k_Z)/2 = 0.012$ µs$^{-1}$ based on the measured decay rates of another individual pentacene-$d_{14}$ molecule.

**Data availability:** The data supporting the findings of this study are available from the corresponding authors upon request.

**Methods references:**

Application to single pentacene molecules in crystalline p-terphenyl. *Phys. Rev. B* **58**, 8997-9017 (1998).

**Acknowledgments:** We thank L. Gross, F. Evers, J. Lupton and R. Huber for discussions and F. Bruckmann, S. Bleher, T. Preis, C. Rohrer, C. Linz and D. Weiss for support. Funding from the ERC Synergy Grant MolDAM (no. 951519) and the Deutsche Forschungsgemeinschaft (DFG, German Research Foundation) through (RE2669/6-2) is gratefully acknowledged.

**Author contributions:** L.S. and J.R. conceived the experiment and L.S., P.S. and R.S. carried them out. L.S., R.S. and J.R. analyzed the experimental results. L.S. and J.R. wrote the manuscript. All authors discussed the results and their interpretation and revised the manuscript.

**Competing interests:** The authors declare no competing interests.

**Additional information:**

**Correspondence and requests for materials** should be addressed to L.S. or J.R.

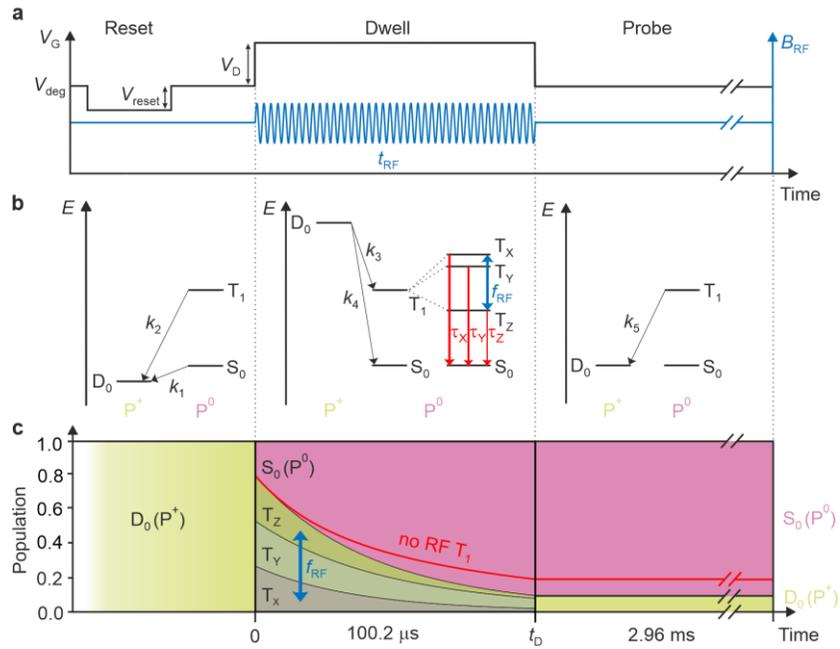

**Extended Data Fig. 1 | Schematic of the pump-probe sequence for the ESR-AFM measurements. a**, Voltage-pulse sequence (black) containing the reset pulse ($V_{\text{reset}} = -1.38$ V, $t_{\text{reset}} = 33.4$ μs) and dwell pulse ($V_{\text{D}} = 2.5$ V, $t_{\text{D}} = 100.2$ μs) applied as $V_{\text{G}}$ to the sample. During the probe interval (2.96 ms), the voltage was set to the middle of the charging hysteresis of $S_0$ and $D_0$ ($V_{\text{deg}}$). An RF pulse (blue) was synchronized with the dwell pulse, having the same duration as the dwell pulse for the measurement of the ESR-AFM spectra. **b**, Many-body picture showing the mutual energetic alignment of the cationic ($P^+$) doublet ground state $D_0$, the neutral ($P^0$) singlet ground state $S_0$ and the neutral triplet excited state $T_1$ during the pump-probe sequence of a. The charge-transfer rates ($k_{1-5}$) were chosen to be much faster than the triplet decay, with $1/k_4$ set to around 4 μs for the ESR-AFM experiments. During the reset pulse, the molecule was brought to $D_0$. The dwell voltage pulse lifted the $D_0$ state above the $T_1$ and $S_0$ state; an electron can tunnel into the molecule forming either the $T_1$ or $S_0$ state, with rate $k_3$ and $k_4$, respectively. Note that rate $k_3$ is approximately a factor 4 larger than $k_4$ due to the reduced tunneling barrier[22]; this ratio depends on the exact tip position, since the two competing

tunneling rates depend on the wave function overlap between tip and the lowest unoccupied and highest occupied orbital of the molecule, respectively. During the dwell pulse, the molecule in $T_1$ can decay back into $S_0$. Two of the triplet states (here $T_X$ and $T_Z$) can be coupled during this time by the RF pulse. If the molecule was still in the triplet state after the dwell pulse, an electron can tunnel out of the molecule charging it, allowing a discrimination of the triplet and singlet state *via* the charged and neutral state, respectively (owing to the reorganization energy, $S_0$ and $D_0$ do not interconvert). c, Populations of the involved states during the pump-probe sequence, with and without RF. Note that with RF, the $T_X$ and $T_Z$ states decay with the average decay rate of the $T_X$ and $T_Z$, assuming a sufficiently strong RF pulse.

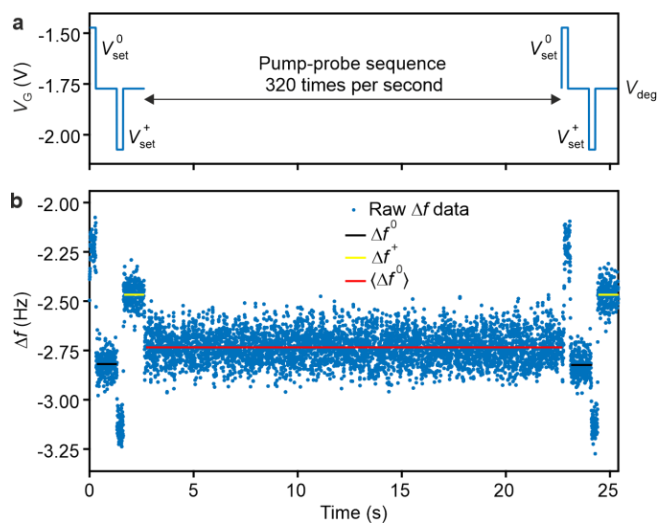

**Extended Data Fig. 2 | Trace showing the raw data acquired for a single ESR data point. a**, Voltage pulse sequence for the acquisition of one data point. At the beginning and end, two voltage pulses were given to neutralize ($V_{set}^0 = V_{deg} + 0.3$ V) and charge ($V_{set}^+ = V_{deg} - 0.3$ V) the molecule, in between which the voltage was set to the middle of the charging hysteresis ($V_{deg}$), here a typical value is shown. During the middle 20 seconds of the data trace the pump-probe sequence shown in Extended Data Fig 1a was repeated 320 times per second. **b**, One of the recorded $\Delta f$ data traces with the pulse sequence shown in a. The frequency shift of the neutral ($\Delta f^0$, black) and charged ($\Delta f^+$, yellow) molecule were extracted as the average over the one second intervals at the beginning and end of the trace. The averaged frequency shift ($\langle \Delta f \rangle$, red) was extracted from the interval during which the pump-probe sequence was turned on. The normalized frequency shift was derived from these three values as $\Delta f_{norm} = \frac{\langle \Delta f \rangle - \Delta f^0}{\Delta f^+ - \Delta f^0}$.

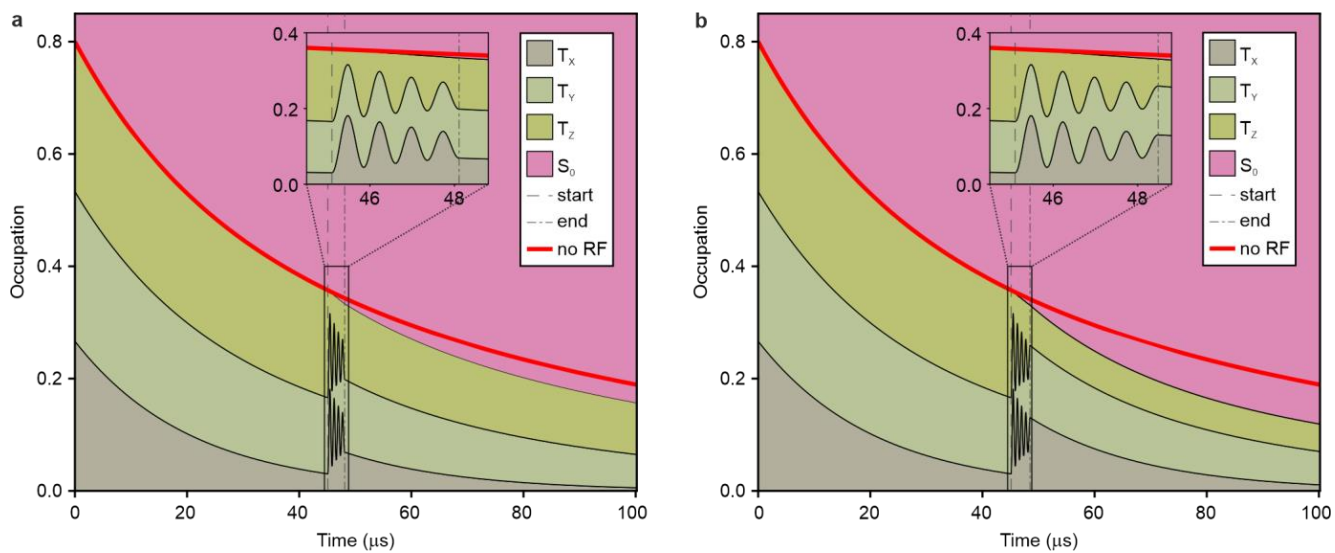

**Extended Data Fig. 3 | Maxwell-Bloch simulations of two data points of the Rabi oscillations.** The occupations of the three triplet states and the singlet state during the dwell voltage pulse ($t_D = 100.2$ μs) are shown. The simulation parameters were chosen similar to those of the measured Rabi oscillations for pentacene-$h_{14}$. At time zero, the beginning of the dwell voltage pulse, it is assumed that the three triplet states are equally occupied; their population summing up to 80% of the total population[22]. During the dwell time, the three triplet states decay with $\tau_X = 21$ μs, $\tau_Y = 67$ μs and $\tau_Z = 136$ μs back to $S_0$. At the starting time $t_S$ an RF pulse is turned on with a frequency matching the $T_X$-$T_Z$ energy splitting. This RF pulse causes coherent oscillations between these two triplet states, as clearly visible in the inset. The population in the $T_X$ and $T_Z$ state at the end of the RF pulse depend, thus, on the duration of the RF pulse. The larger the population in the fastest decaying $T_X$ state, the lower the triplet population at the end of the dwell pulse. This is exemplified by the simulations shown in **a** and **b** with RF pulse durations corresponding to 4 and 4.5 Rabi-oscillation periods, respectively (Rabi frequency: 1.33 MHz, decay constant of the oscillations: 2.2 μs).

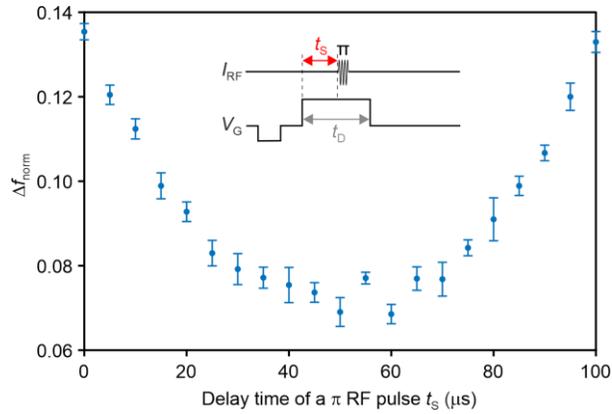

**Extended Data Fig. 4 | The effect of the delay time of a π RF pulse on the normalized frequency shift.** The population difference of the $T_X$ and $T_Z$ state can be driven with a π RF pulse from the $T_Z$ to the $T_X$ state, after which this population can decay with the fast decay rate of the $T_X$ state. At short delay times, only a small signal is observed, due to the equal initial occupation of the three triplet states. The population difference increases with delay time, initially increasing the detected signal ($\Delta f_{\text{norm}}$ drops). In case of long delay times, the signal reduces again ($\Delta f_{\text{norm}}$ rises), because the remaining time during the dwell voltage pulse becomes too short for the $T_X$ population to decay. This data is measured for the same pentacene-$d_{14}$ molecule as in Fig 3b. The inset shows the used pump-probe pulse scheme with a π RF pulse (duration: 0.29 μs) after a variable delay time ($t_S$) with respect to the start of the dwell voltage pulse (duration: $t_D = 100.2$ μs).